\documentclass[twoside]{article}

\usepackage{graphicx}

\usepackage{natbib}
\bibpunct{(}{)}{;}{a}{}{,}
\newlength{\figurewidth}
\renewcommand{\sectionmark}[1]{\markboth{\textsc{M. E. J.
Newman}}{\textsc{Betweenness and random walks}}}
\renewcommand{\subsectionmark}[1]{\markboth{\textsc{M. E. J.
Newman}}{\textsc{Betweenness and random walks}}}

\newcommand{\Ord}{\mathrm{O}}

\newcommand{\half}{\mbox{$\frac12$}}

\newcommand{\eref}[1]{(\ref{#1})}
\newcommand{\etal}{{\it{}et~al.}}
\newcommand{\defn}{\textit}
\newcommand{\citey}{\citeyearpar}

\newcommand{\vI}{\mathbf{I}}
\newcommand{\vM}{\mathbf{M}}
\newcommand{\vV}{\mathbf{V}}
\newcommand{\vD}{\mathbf{D}}
\newcommand{\vA}{\mathbf{A}}
\newcommand{\vT}{\mathbf{T}}

\newcommand{\vs}{\mathbf{s}}

\newcommand{\captionfonts}{\small}
\makeatletter
\long\def\@makecaption#1#2{%
  \vskip\abovecaptionskip
  \sbox\@tempboxa{{\captionfonts #1: #2}}%
  \ifdim \wd\@tempboxa >\hsize
    {\captionfonts #1: #2\par}
  \else
    \hbox to\hsize{\hfil\box\@tempboxa\hfil}%
  \fi
  \vskip\belowcaptionskip}
\makeatother

\begin{document}

\title{A measure of betweenness centrality\\
based on random walks}
\author{M. E. J. Newman\\
\\
\textit{\normalsize Department of Physics and
Center for the Study of Complex Systems,}\\
\textit{\normalsize University of Michigan, Ann Arbor, MI 48109--1120}}
\date{}
\maketitle

\begin{abstract}
Betweenness is a measure of the centrality of a node in a network, and is
normally calculated as the fraction of shortest paths between node pairs
that pass through the node of interest.  Betweenness is, in some sense, a
measure of the influence a node has over the spread of information through
the network.  By counting only shortest paths, however, the conventional
definition implicitly assumes that information spreads only along those
shortest paths.  Here we propose a betweenness measure that relaxes this
assumption, including contributions from essentially all paths between
nodes, not just the shortest, although it still gives more weight to short
paths.  The measure is based on random walks, counting how often a node is
traversed by a random walk between two other nodes.  We show how our
measure can be calculated using matrix methods, and give some examples of
its application to particular networks.
\end{abstract}

\section{Introduction}
\label{intro}
Over the years network researchers have introduced a large number of
\defn{centrality indices}, measures of the varying importance of the
vertices in a network according to one criterion or another
\citep{WF94,Scott00}.  These indices have proved of great value in the
analysis and understanding of the roles played by actors in social
networks,\footnote{``Actor'' is the generic term used by sociologists to
refer to a node in a social network.} as well as by vertices in networks of
other types, including citation networks, computer networks, and biological
networks.  Perhaps the simplest centrality measure is \defn{degree}, which
is the number of edges incident on a vertex in a network---the number of
ties an actor has in social network parlance.  Degree is a measure in some
sense of the popularity of an actor.  A more sophisticated centrality
measure is \defn{closeness}, which is the mean geodesic
(i.e.,~shortest-path) distance between a vertex and all other vertices
reachable from it.\footnote{Some define closeness to be the reciprocal of
this quantity, but either way the information communicated by the measure
is the same.}  Closeness can be regarded as a measure of how long it will
take information to spread from a given vertex to others in the network.

Another important class of centrality measures is the class of betweenness
measures.  Betweenness, as one might guess, is a measure of the extent to
which a vertex lies on the paths between others.  The simplest and most
widely used betweenness measure is that of
Freeman~\citey{Freeman77,Freeman79}, usually called simply
\defn{betweenness}.  (Where necessary, to distinguish this measure from
other betweenness measures considered in this paper, we will refer to it as
\defn{shortest-path betweenness}.)  The betweenness of a vertex~$i$ is
defined to be the fraction of shortest paths between pairs of vertices in a
network that pass through~$i$.  If, as is frequently the case, there is
more than one shortest path between a given pair of vertices, then each
such path is given equal weight such that the weights sum to unity.  To be
precise, suppose that $g^{(st)}_i$ is the number of geodesic paths from
vertex~$s$ to vertex~$t$ that pass through~$i$, and suppose that $n_{st}$
is the total number of geodesic paths from $s$ to~$t$.  Then the
betweenness of vertex~$i$ is
\begin{equation}
b_i = {\sum_{s<t} g^{(st)}_i/n_{st}\over\frac12 n(n-1)},
\end{equation}
where $n$ is the total number of vertices in the
network.\footnote{Alternatively, $b_i$ may be normalized by dividing by its
maximum possible value, which it achieves for a ``star graph'' in which one
central vertex is connected to every other by a single edge
\citep{Freeman79}.}  We may, or may not, according to taste, consider the
end-points of a path to fall on that path; the choice makes only the
difference of an additive constant in the values for $b_i$.  In this paper
we will generally include the end-points.

Betweenness centrality can be regarded as a measure of the extent to which
an actor has control over information flowing between others.  In a network
in which flow is entirely or at least mostly along geodesic paths, the
betweenness of a vertex measures how much flow will pass through that
particular vertex.  Betweenness can be calculated for all vertices in time
$\Ord(mn)$ for a network with $m$ edges and $n$ vertices
\citep{Newman01c,Brandes01}.

In most networks however, information (or anything else) does not flow only
along geodesic paths \citep{SZ89,FBW91}.  News or a rumor or a message or a
fad does not know the ideal route to take to get from one place to another;
more likely it wanders around more randomly, encountering who it will.  And
even in a case such as the famous small-world experiment of Milgram
(\citeyear{Milgram67}; \citealt{TM69}), or its modern-day equivalent
\citep{DMW03}, in which participants are explicitly instructed to get a
message to a target by the most direct route possible, there is no evidence
that people are especially successful in this task.  Thus we would imagine
that in most cases a realistic betweenness measure should include
non-geodesic paths in addition to geodesic ones.

\begin{figure}
\begin{center}
\resizebox{\columnwidth}{!}{\includegraphics{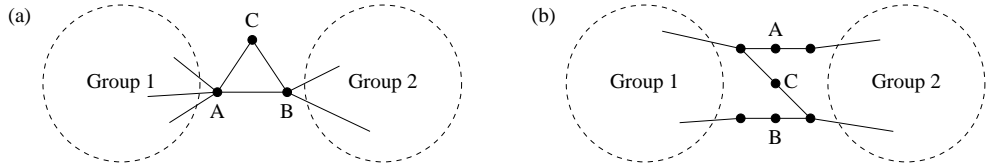}}
\end{center}
\caption{(a)~Vertices A and B will have high (shortest-path) betweenness in
this configuration, while vertex~C will not.  (b)~In calculations of flow
betweenness, vertices A and B in this configuration will get high scores
while vertex~C will not.}
\label{problem}
\end{figure}

Furthermore, by giving all the weight to the geodesic paths, and none to
any other paths, no matter how closely competitive they are, the
shortest-path betweenness measure can produce some odd effects.  Consider
the network sketched in Fig.~\ref{problem}a, for instance, in which two
large groups are bridged by connections among just a few of their members.
Vertices~A and~B will certainly get high betweenness scores in this case,
since all shortest paths between the two communities must pass through
them.  Vertex~C on the other will hand get a low score, since none of those
shortest paths pass through it, taking instead the direct route from A
to~B.  It is plausible however that in many real-world situations C would
have quite a significant role to play in information flows.  Certainly it
is possible for information to flow between two individuals via a third
mutual acquaintance, even when the two individuals in question are
themselves well acquainted.

To address these problems, Freeman~\etal~\citey{FBW91} suggested a more
sophisticated betweenness measure, usually known as \defn{flow
betweenness}, that includes contributions from some non-geodesic paths.
Flow betweenness is based on the idea of maximum flow.  Imagine each edge
in a network as a pipe that can carry a unit flow of some fluid.  We can
ask what the maximum possible flow then is between a given source
vertex~$s$ and target vertex~$t$ through these pipes.  In general the
answer is that more than a single unit of flow can be carried between
source and target by making simultaneous use of several different paths
through the network.  The flow betweenness of a vertex~$i$ is defined as
the amount of flow through vertex~$i$ when the maximum flow is transmitted
from $s$ to~$t$, averaged over all $s$ and~$t$.\footnote{Technically, this
definition is not unique, because there need not be a unique solution to
the flow problem.  To get around this difficulty, Freeman~\etal\ define
their betweenness measure as the the maximum possible flow through~$i$ over
all possible solutions to the $st$ maximum flow problem, averaged over all
$s$ and~$t$.}  Maximum flow from a given $s$ to all reachable targets~$t$
can be calculated in worst-case time $\Ord(m^2)$ using, for instance, the
augmenting path algorithm \citep{AMO93}, and hence the flow betweenness for
all vertices can be calculated in time $\Ord(m^2n)$.\footnote{One can do
somewhat better, particularly on networks like those discussed here in
which all edges have the same capacity, by using more advanced algorithms.
See Ahuja~\etal~\citey{AMO93}.}

In practical terms, one can think of flow betweenness as measuring the
betweenness of vertices in a network in which a maximal amount of
information is continuously pumped between all sources and targets.
Necessarily, that information still needs to ``know'' the ideal route (or
one of the ideal routes) from each source to each target, in order to
realize the maximum flow.  Although the flow betweenness does take account
of paths other than the shortest path (and indeed need not take account of
the shortest path at all), this still seems an unrealistic definition for
many practical situations: flow betweenness suffers from some of the same
drawbacks as shortest-path betweenness, in that it is often the case that
flow does not take any sort of ideal path from source to target, be it the
shortest path, the maximum flow path, or another kind of ideal path.

Moreover, like the shortest-path measure, flow betweenness can give
counterintuitive results in some cases.  Consider for example the network
sketched in Fig.~\ref{problem}b, which again has two large groups joined by
a few contacts.  In this case, the maximum flow from one group to the other
is clearly limited to two units, one unit flowing through each of vertices
A and~B.  Vertex~C will, in this case, get a low betweenness score, even
though the path through C may be as short or shorter than that through A
or~B.  Once again, it is plausible that in practical situations C would
actually play quite a significant role.

In this paper, therefore, we propose a new betweenness measure, which might
be called \defn{random-walk betweenness}.  Roughly speaking, the
random-walk betweenness of a vertex~$i$ is equal to the number of times
that a random walk starting at~$s$ and ending at~$t$ passes through~$i$
along the way, averaged over all $s$ and~$t$.  This measure is appropriate
to a network in which information wanders about essentially at random until
it finds its target, and it includes contributions from many paths that are
not optimal in any sense, although shorter paths still tend to count for
more than longer ones since it is unlikely that a random walk becomes very
long without finding the target.  In some sense, our random-walk
betweenness and the shortest-path betweenness of Freeman~\citey{Freeman77}
are at opposite ends of a spectrum of possibilities, one end representing
information that has no idea of where it is going and the other information
that knows precisely where it is going.  Some real-world situations may
mimic these extremes while others, such as perhaps the small-world
experiment, fall somewhere in between.  In the latter case it may be of use
to compare the predictions of the two measures to see how and by how much
they differ: if they differ little, then either is a reasonable metric by
which to characterize the system; if they differ by a lot, then we may need
to know more about the particular mode of information propagation in the
network to make meaningful judgments about betweenness of vertices.

Our random-walk betweenness can, as we will show, be calculated for all
vertices in a network in worst-case time $\Ord((m+n)n^2)$ using matrix
methods, making it comparable in its computational demands with flow
betweenness.

Some other centrality measures based on random walks merit a mention in
this context, although none of them are betweenness measures.  Bonacich's
\defn{power centrality} \citep{Bonacich87} can be derived in a number of
ways, but one way of looking at it is in terms of random walks that have a
fixed probability $\beta$ of dying per step.  The power centrality of
vertex~$i$ is the expected number of times such a walk passes through~$i$,
averaged over all possible starting points for the walk.  The
\defn{random-walk centrality} introduced recently by Noh and
Rieger~\citey{NR03} is a measure of the speed with which randomly walking
messages reach a vertex from elsewhere in the network---a sort of
random-walk version of closeness centrality.  The
\defn{information centrality} of Stephenson and Zelen~\citey{SZ89} is
another closeness measure, which bears some similarity to that of Noh and
Rieger.  In essence it measures the harmonic mean length of paths ending at
a vertex~$i$, which is smaller if $i$ has many short paths connecting it to
other vertices.

The outline of this paper is as follows.  In Sec.~\ref{rw} we define in
detail our random-walk betweenness and show how it is calculated.  We
introduce the measure first using an analogy to the flow of electrical
current in a circuit, and then show that this is equivalent also to the
flow of a random walk.  In Sec.~\ref{examples} we give a number of examples
of applications of our measure, first to networks artificially designed to
pose a challenge for the calculation of betweenness, and then to various
real-world social networks, including a collaboration network of
scientists, a network of sexual contacts, and Pagdett's network of
intermarriages between prominent families in 15th century Florence.  In
Sec.~\ref{concs} we give our conclusions.

\section{Random-walk betweenness}
\label{rw}
In this section we give the definition of our random-walk betweenness
measure and derive matrix expressions that allow it to be calculated
rapidly using a computer.  For pedagogical purposes, we will take a
slightly circuitous route in developing our ideas.  We start by introducing
a definition of our betweenness measure that does not use random walks but
instead is based on current flow in electrical circuits.  This definition
is simple and intuitive and makes for easy calculations.  Later we
introduce the random-walk definition of our measure and prove that the two
definitions are the same.  The developments of this section follow similar
lines to a previous presentation we have given on methods for hierarchical
clustering
\citep{NG04}.

\begin{figure}[b]
\begin{center}
\resizebox{8cm}{!}{\includegraphics{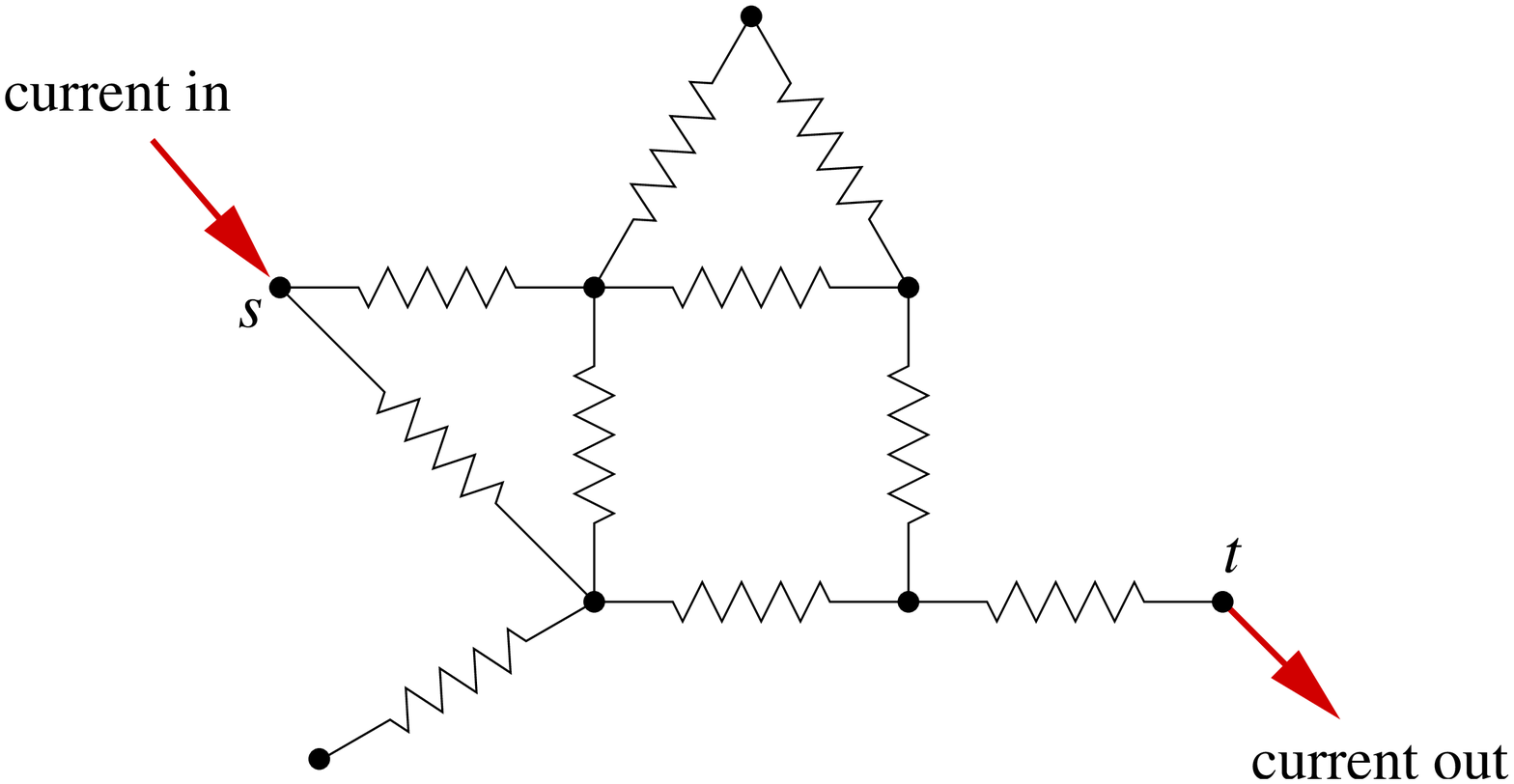}}
\end{center}
\caption{An electrical circuit as discussed in the text, in which the edges
of a network have been replaced by identical unit resistors, and unit
current is injected at vertex~$s$ and removed at vertex~$t$.  The
betweenness of a vertex is defined to be equal to the current flowing
through that vertex averaged over all $s$ and~$t$.}
\label{resistor}
\end{figure}

\subsection{A current flow analogy}
\label{current}
Consider, then, an electrical circuit created by placing a unit resistance
on every edge of the network of interest, as shown in Fig.~\ref{resistor}.
One unit of current is injected into the network at a source vertex~$s$ and
one unit extracted at a target vertex~$t$, so that current in the network
as a whole is conserved.  We now define the \defn{current-flow betweenness}
of a vertex~$i$ to be the amount of current that flows through~$i$ in this
setup, averaged over all $s$ and~$t$.

Let $V_i$ be the voltage at vertex~$i$ in the network, measured relative to
any convenient point.  Kirchhoff's law of current conservation states that
the total current flow into or out of any vertex is zero, which implies
that the voltages satisfy the equations
\begin{equation}
\sum_j A_{ij}(V_i-V_j) = \delta_{is} - \delta_{it},
\label{kirchhoff1}
\end{equation}
for all~$i$, where $A_{ij}$ is an element of the adjacency matrix thus:
\begin{equation}
A_{ij} = \biggl\lbrace\begin{array}{rl}
           1 & \qquad\mbox{if there is an edge between $i$ and $j$,}\\
           0 & \qquad\mbox{otherwise,}
         \end{array}
\label{adjacency}
\end{equation}
and $\delta_{ij}$ is the Kronecker~$\delta$:
\begin{equation}
\delta_{ij} = \biggl\lbrace\begin{array}{rl}
                1 & \qquad\mbox{if $i=j$,}\\
                0 & \qquad\mbox{otherwise.}
              \end{array}
\end{equation}
Noting that $\sum_j A_{ij} = k_i$, the degree of vertex~$i$, we can write
Eq.~\eref{kirchhoff1} in matrix form as
\begin{equation}
(\vD - \vA)\cdot\vV = \vs,
\label{kirchhoff2}
\end{equation}
where $\vD$ is the diagonal matrix with elements $D_{ii}=k_i$ and the
source vector $\vs$ has elements
\begin{equation}
s_i = \Biggl\lbrace\begin{array}{rl}
        +1 & \qquad\mbox{for $i=s$,}\\
        -1 & \qquad\mbox{for $i=t$,}\\
        0  & \qquad\mbox{otherwise.}
      \end{array}
\label{defssi}
\end{equation}

We cannot simply invert the matrix $\vD-\vA$ to get the voltage
vector~$\vV$, because the matrix (which is called the \defn{graph
Laplacian}) is singular: the vector $\vV=(1,1,1,\ldots)$ is always an
eigenvector with eigenvalue zero because voltage is arbitrary to within an
additive constant, and since the determinant is the product of the
eigenvalues, it follows that the determinant is always zero.
Mathematically, this says that one of the equations in our system of $n$
equations is redundant.  (Physically, it is telling us that current is
conserved.)  To fix the problem we need only choose one equation, any
equation, and remove it from the system, to get a matrix we can invert.
This operation is made most simple if we simultaneously choose to measure
our voltages relative to the corresponding vertex.  Thus, let us measure
voltages relative to some vertex~$v$, and remove the $v$th equation, which
means removing the $v$th row of $\vD-\vA$.  Since $V_v=0$, we can also
remove the $v$th column, giving a square $(n-1)\times(n-1)$ matrix, which
we denote $\vD_v-\vA_v$.  Then
\begin{equation}
\vV = (\vD_v-\vA_v)^{-1}\cdot\vs.
\label{vv1}
\end{equation}

The voltage of the one missing vertex~$v$ is, by definition, zero.  To
represent this, let us now add a $v$th row and column back into
$(\vD_v-\vA_v)^{-1}$ with values all equal to zero.  The resulting matrix
we will denote~$\vT$.  Then, using Eq.~\eref{defssi}, the voltage at
vertex~$i$ for source~$s$ and target~$t$ is given in terms of the elements
of $\vT$ by
\begin{equation}
V^{(st)}_i = T_{is} - T_{it}.
\end{equation}

The current flowing through the $i$th vertex is given by a half of the sum
of the absolute values of the currents flowing along the edges incident on
that vertex:
\begin{equation}
I^{(st)}_i = \half \sum_j A_{ij} |V^{(st)}_i - V^{(st)}_j|
           = \half \sum_j A_{ij} | T_{is} - T_{it} - T_{js} + T_{jt} |,
             \quad\mbox{for $i\ne s,t$.}
\label{ist}
\end{equation}
As noted, this expression does not work for the source and target vertices,
for which one also has to take account of the current injected and removed
from the network, but these vertices necessarily have a current flow of
exactly one unit, so one can simply write
\begin{equation}
I^{(st)}_s = 1,\qquad I^{(st)}_t = 1.
\label{endpoints}
\end{equation}
(Alternatively, if one is adopting the convention under which the
end-points of a path are not considered part of that path, then one should
set these two currents to 0 rather than~1.)  Then the betweenness is the
average of the current flow over all source-target pairs:
\begin{equation}
b_i = {\sum_{s<t} I^{(st)}_i\over\frac12 n(n-1)}.
\label{defsbi}
\end{equation}
If the network has more than one component, then the procedure described
here should be repeated separately for each component.

The inversion of the matrix takes time $\Ord(n^3)$, while the evaluation of
Eq.~\eref{defsbi} takes time $\Ord(mn)$ for each vertex or $\Ord(mn^2)$ for
all of them.  Thus the total running time to calculate the current-flow
betweenness for all vertices is $\Ord((m+n)n^2)$, or $\Ord(n^3)$ on a
sparse graph.  This is comparable with the time for calculation of the flow
betweenness, although slower than the fastest algorithms for shortest-path
betweenness.  In our experience, the calculation is tractable for networks
up to about $10\,000$ vertices using typical desktop computing resources
available at the time of writing.\footnote{In fact, as computer hardware
stands at present, one is more likely to run out of memory than time, the
memory requirements for matrix inversion being $\Ord(n^2)$, which means a
gigabyte or more for a $10\,000\times10\,000$ matrix.  It is possible that
larger systems could be tackled using specialized sparse-matrix inversion
methods, although there is a penalty in running time to be paid for doing
this.}

This current-flow betweenness measure seems an intuitively reasonable one.
Current will flow along all paths from source to target, but more along
shorter than longer ones, the shorter ones offering less resistance than
the longer.  And vertices that lie on no path from source to target, those
that sit in a \textit{cul-de-sac} off to the side of the network, get a
betweenness of zero, which is also sensible.  However, there is no special
reason to believe that the flow of electrical current has anything to do
with processes in non-electrical networks, such as social networks.  In the
following section, therefore, we introduce our random-walk betweenness,
whose definition, we believe, \emph{is} relevant for social networks and
other types of networks also, and we show that it is in fact numerically
equal to the current-flow betweenness.

\subsection{Random walks}
Imagine a ``message,'' which could be information of almost any kind, that
originates at a source vertex~$s$ on a network.  The message is intended
for some target~$t$, but the message, or those passing it, have no idea
where $t$ is, so the message simply gets passed around at random until it
finds itself at~$t$.  Thus, on each step of its travels, the message moves
from its current position on the network to one of the adjacent vertices,
chosen uniformly at random from the possibilities.  This is a \defn{random
walk}.

With the important proviso discussed in the following paragraph, let us
define a betweenness measure for a vertex~$i$ that is equal to the number
of times that the message passes through $i$ on its journey, averaged over
a large number of trials of the random walk.  The full random-walk
betweenness of vertex~$i$ will then be this value averaged over all
possible source/target pairs $s,t$.

There is one small, but important, further technical point.  It would be
perfectly possible for a vertex to accrue a high betweenness score if a
random walk were simply to walk back and forth through that vertex many
times, without actually going anywhere.  This situation does not correspond
to our intuition of what it means to have high betweenness, and indeed if
we count walks in this way our betweenness measure is found to give mostly
useless results.  So instead, we define the betweenness of vertex~$i$ to be
the \emph{net} number of times a walk passes through~$i$.  By ``net'' we
mean that if a walk passes through a vertex and then later passes back
through it in the opposite direction, the two cancel out and there is no
contribution to the betweenness.  Furthermore, if, when averaged over many
possible realizations of the walk in question, we find that the walk is
equally likely to pass in either direction through a vertex, then again the
two directions cancel.  How we allow for these cancellations in practice
will become clear shortly.

Consider then an \defn{absorbing random walk}, a walk that starts at
vertex~$s$ and makes random moves around the network until it finds itself
at vertex~$t$ and then stops.  If at some point in this walk we find
ourselves at vertex~$j$, then the probability that we will find ourselves
at~$i$ on the next step is given by the matrix element
\begin{equation}
M_{ij} = {A_{ij}\over k_j},\qquad\mbox{for $j\ne t$,}
\label{defsm}
\end{equation}
where once again $A_{ij}$ is an element of the adjacency matrix,
Eq.~\eref{adjacency}, and $k_j=\sum_i A_{ij}$ is the degree of vertex~$j$.
In matrix notation, we can write $\vM=\vA\cdot\vD^{-1}$, where $\vD$ is, as
before, the diagonal matrix with elements $D_{ii}=k_i$.

The only exception to Eq.~\eref{defsm}, as noted, is for $j=t$; since this
is an absorbing random walk, we never leave $t$ once we get there (a
behavior sometimes jokingly called the ``Hotel California effect''), so
$M_{it}=0$ for all~$i$.  Alternatively, we can simply remove row~$t$ from
the matrix altogether.  We can also remove column~$t$ without affecting
transitions between any other vertices.  Let us denote by
$\vM_t=\vA_t\cdot\vD_t^{-1}$ the matrix with these elements removed, and
similarly for $\vA_t$ and~$\vD_t$.

Now for a walk starting at~$s$, the probability that we find ourselves at
vertex~$j$ after $r$ steps is given by $[\vM_t^r]_{js}$, and the
probability that we then take a step to an adjacent vertex~$i$ is $k_j^{-1}
[\vM_t^r]_{js}$.  Summing over all values of $r$ from 0 to~$\infty$, the
total number of times we go from $j$ to $i$, averaged over all possible
walks, is $k_j^{-1} [(\vI-\vM_t)^{-1}]_{js}$.  In matrix notation we can
write this as an element of the vector
\begin{equation}
\vV = \vD_t^{-1}\cdot(\vI-\vM_t)^{-1}\cdot\vs
    = (\vD_t-\vA_t)^{-1}\cdot\vs,
\label{rwres}
\end{equation}
where $\vs$ is defined as before---see Eq.~\eref{defssi}.  (The element
$s_t=-1$ is not strictly necessary---we could give $s_t$ any value we like,
since row $t$ is removed from the equations anyway.  We make this
particular choice in order to demonstrate that our random-walk betweenness
is the same as the current-flow betweenness.)  Clearly this equation is
precisely the same as Eq.~\eref{vv1}, for the particular choice $v=t$.

Now the net flow of the random walk along the edge from $j$ to $i$ is given
by the absolute difference $|V_i-V_j|$ and the net flow through vertex~$i$
is a half the sum of the flows on the incident edges, just as in
Eq.~\eref{ist}.  The rest of the derivation follows through as before, and
the final net flow of random walks through vertex~$i$ is given by
Eq.~\eref{defsbi}.  (Although Eq.~\eref{rwres} was derived for the
particular case in which the $t$th row and column are removed from the
matrix $\vD-\vA$, the developments of Sec.~\ref{current} show that the same
result can be derived by removing any row and column.)

To summarize, the prescription for calculating random-walk betweenness,
which is the expected net number of times a random walk passes through
vertex~$i$ on its way from a source~$s$ to a target~$t$, averaged over all
$s$ and~$t$, is as follows for each separate component of the graph of
interest.
\begin{enumerate}
\item Construct the matrix $\vD-\vA$, where $\vD$ is the diagonal matrix of
vertex degrees and $\vA$ is the adjacency matrix.
\item Remove any single row, and the corresponding column.  For example,
one could remove the last row and column.
\item Invert the resulting matrix and then add back in a new row and column
consisting of all zeros in the position from which the row and column were
previously removed (e.g.,~the last row and column).  Call the resulting
matrix~$\vT$, with elements $T_{ij}$.
\item Calculate the betweenness from Eq.~\eref{defsbi}, using the values of
$I_i$ from Eqs.~\eref{ist} and~\eref{endpoints}.
\end{enumerate}

\section{Examples and applications}
\label{examples}
In this section we give a number of examples to illustrate the properties
and use of our random-walk betweenness measure in the analysis of network
data.

\subsection{Simple examples}
\begin{figure}[b]
\begin{center}
\resizebox{\columnwidth}{!}{\includegraphics{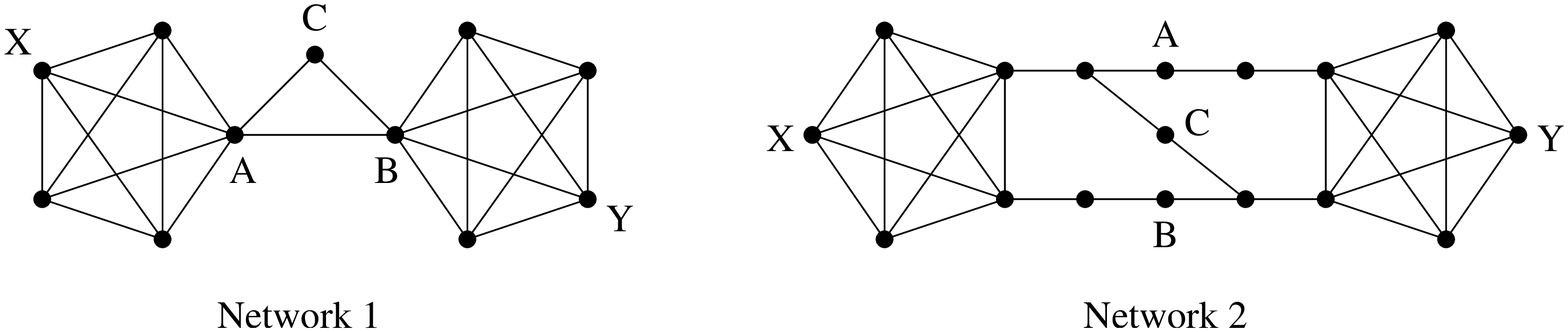}}
\end{center}
\caption{Example networks of the types sketched in Fig.~\ref{problem}, with
the groups represented by completely connected graphs of five vertices
each.}
\label{fivefive}
\end{figure}

We start off by presenting some cases with which previous betweenness
measures have difficulties, but for which, as we now show, our new measure
works well.  Our examples are based on the graphs sketched in
Fig.~\ref{problem}, with the two groups consisting of complete graphs of
five vertices each, as depicted in Fig.~\ref{fivefive}.  We show figures
for the betweenness scores in Table~\ref{tab55}.

\begin{table}[t]
\begin{center}
\setlength{\tabcolsep}{8pt}
\begin{tabular}{ll|ccc}
 & & \multicolumn{3}{c}{betweenness measure}  \\
\cline{3-5}
\multicolumn{2}{c|}{network} & shortest-path  &           flow & random-walk \\
\hline
Network 1: & vertices A \& B &       $0.636$  &        $0.631$ &     $0.670$ \\
           & vertex C        & \fbox{$0.200$} &        $0.282$ &     $0.333$ \\
           & vertices X \& Y &       $0.200$  &        $0.068$ &     $0.269$ \\
\hline
Network 2: & vertices A \& B &       $0.265$  &        $0.269$ &     $0.321$ \\
           & vertex C        &       $0.243$  & \fbox{$0.004$} &     $0.267$ \\
           & vertices X \& Y &       $0.125$  &        $0.024$ &     $0.194$ \\
\end{tabular}
\end{center}
\caption{Betweenness values calculated using shortest-path, flow, and
random-walk measures for the two networks of Fig.~\ref{fivefive}.  In each
network, we intuitively expect vertex~C to have betweenness lower than that
of vertices A and~B, but higher than that of vertices X and~Y.  The
shortest-path and flow betweenness measures each fail to do this for one of
these two challenging networks (numbers in boxes).  Our random-walk measure
on the other hand orders the vertices correctly in each case.}
\label{tab55}
\end{table}

As the table shows, the shortest path betweenness fails to give a higher
score to vertex~C in the first network than to any of the other vertices
within the two communities, while flow betweenness has the same problem
with vertex~C in the second network.  In both cases, by contrast, our
random-walk betweenness gives vertex~C a distinctly higher score,
reflecting our intuition that this vertex has a higher centrality in both
of the networks.

\subsection{Correlation with other measures}
Shortest-path betweenness is known to be strongly correlated with vertex
degree in most networks \citep{Nakao90,GOKK03}, and it has been argued that
this makes it a less useful measure.  If the two are strongly correlated,
then what is the point of going to the effort of calculating betweenness,
when degree is almost the same and much easier to calculate?  The answer is
that there are usually a small number of vertices in a network for which
betweenness and degree are very different, and betweenness is useful
precisely in identifying these vertices.

\begin{figure}
\begin{center}
\resizebox{12cm}{!}{\includegraphics{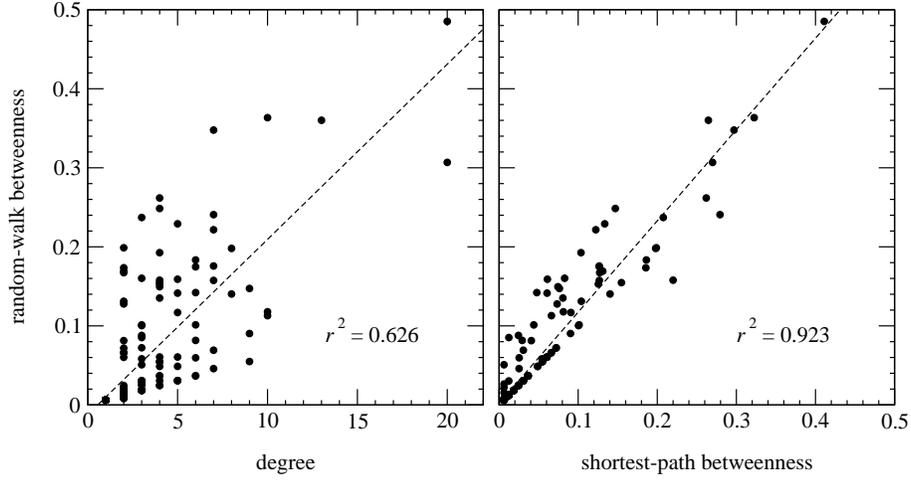}}
\end{center}
\caption{Scatter plots of the random-walk betweenness of vertices in the
sexual contact network of Fig.~\ref{potterat}, against vertex degree (left)
and standard shortest-path betweenness (right).  The dotted lines indicate
the best linear fits in each case, which have the correlation coefficients
indicated.}
\label{correlations}
\end{figure}

We can look at a similar question for our random-walk measure.  In
Fig.~\ref{correlations}, for example, we show scatter plots of random-walk
betweenness vs.\ (a)~degree and (b)~shortest-path betweenness for the
actors in a network of sexual contacts drawn from the study of
Potterat~\etal~\citey{Potterat02}.  As the figure shows, the random-walk
betweenness is moderately highly correlated with degree ($r^2=0.626$) and
very highly correlated with shortest-path betweenness ($r^2=0.923$).  Thus,
in general, vertices with higher degree or higher shortest-path betweenness
tend also to have higher random-walk betweenness.  However, this
observation misses the real point of interest, which is that there are a
few vertices that have random-walk betweenness values quite different from
their scores on the other two measures.

\begin{figure}
\begin{center}
\resizebox{\columnwidth}{!}{\includegraphics{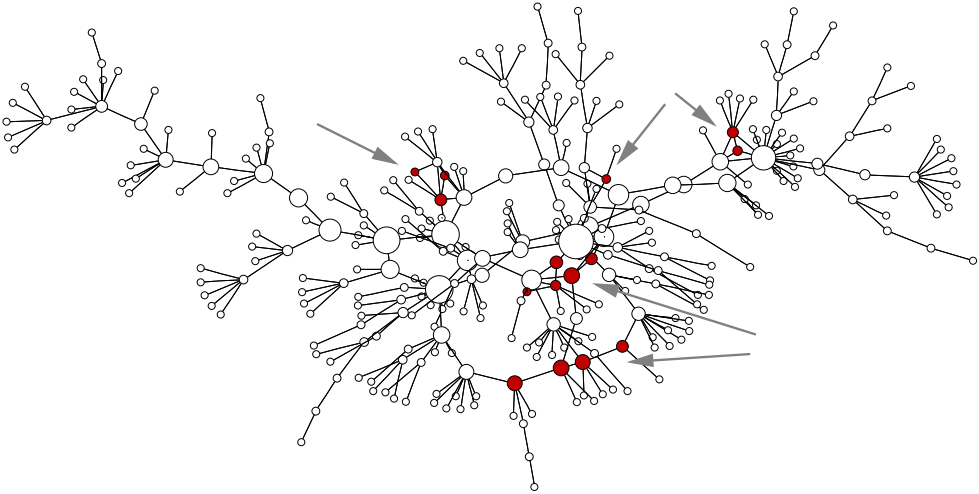}}
\end{center}
\caption{The largest component of a network of sexual contacts between
high-risk actors in the city of Colorado Springs, CO, as reconstructed by
Potterat~\etal~\citey{Potterat02}.  The size of the vertices increases
linearly with their random-walk betweenness, as defined in this paper.  The
highlighted vertices (also indicated by the arrows) are those for which the
random-walk betweenness is substantially greater than shortest-path
betweenness (a factor of two or more).}
\label{potterat}
\end{figure}

In Fig.~\ref{potterat} we show a picture of the network in question, in
which we have drawn each the vertices with a size indicating their
random-walk betweenness score.  It is immediately clear that some, but not
all, of the high-degree vertices have high random-walk betweenness.
Furthermore, we have highlighted the vertices in the network for which the
random-walk betweenness is more than twice their shortest-path
betweenness---these are vertices which the shortest-path measure misses
because, although they lie on many paths between others, they don't lie on
many \emph{shortest} paths.

The primary reason for the study of networks of sexual contacts is to
improve our understanding of the propagation and control of sexually
transmitted diseases.  Certainly there is no reason to suppose that
diseases always know precisely where they are going and spread along the
shortest path to some ``target'' victim.  A random-walk model of disease
spread is probably a more reasonable representation of what actually
happens, in which case the highlighted nodes in Fig.~\ref{potterat} are
nodes that are likely to be responsible for transmission of the disease to
others, but which would be missed if we evaluated the centralities using
standard shortest-path-based methods.

\begin{figure}[t]
\begin{center}
\resizebox{6cm}{!}{\includegraphics{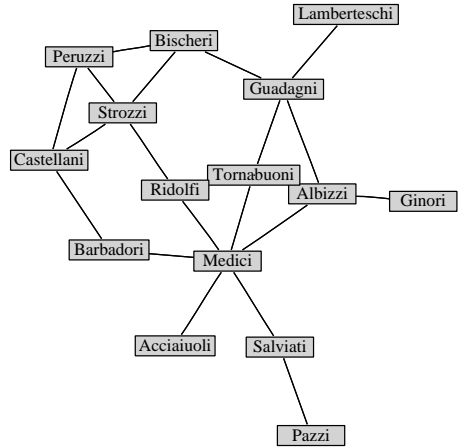}}
\end{center}
\caption{The network of intermarriage relations between the 15th century
Florentine families studied by Padgett and Ansell~\citey{PA93}.  One
family, Pucci, which had no marriage ties with others, is omitted from the
picture.}
\label{padgett}
\end{figure}

\subsection{Example applications}
We now give two brief examples of applications of our betweenness measure
to previously studied networks.  First, we look at Padgett's famous network
of intermarriages between prominent families in early 15th century Florence
\citep{PA93}, depicted in Fig.~\ref{padgett}.  In Table~\ref{families}, we
rank the fifteen families by their random-walk betweenness, finding that
the Medici come out well ahead of the competition, and in particular, they
easily best their arch-rivals, the Strozzi.  It is suggested that that it
was in part the Medici's skillful manipulation of this marriage network
that led to their eventual dominance of the Florentine political landscape.

\begin{table}[t]
\begin{center}
\setlength{\tabcolsep}{12pt}
\renewcommand{\baselinestretch}{1}
\begin{tabular}{l|c||l|c}
family       & betweenness & family       & betweenness \\
\hline
Medici       & $0.652420$  & Barbadori    & $0.269363$  \\
Guadagni     & $0.451309$  & Salviati     & $0.257143$  \\
Albizzi      & $0.362961$  & Peruzzi      & $0.245624$  \\
Strozzi      & $0.333302$  & Pazzi        & $0.133333$  \\
Ridolfi      & $0.317014$  & Lamberteschi & $0.133333$  \\
Bischeri     & $0.314018$  & Ginori       & $0.133333$  \\
Tornabuoni   & $0.306102$  & Acciaiuoli   & $0.133333$  \\
Castellani   & $0.284705$  &              &             \\
\end{tabular}
\end{center}
\caption{The random-walk betweenness scores of the fifteen families in
the network of Fig.~\ref{padgett}.}
\label{families}
\end{table}

\begin{figure}[t]
\begin{center}
\resizebox{\figurewidth}{!}{\includegraphics{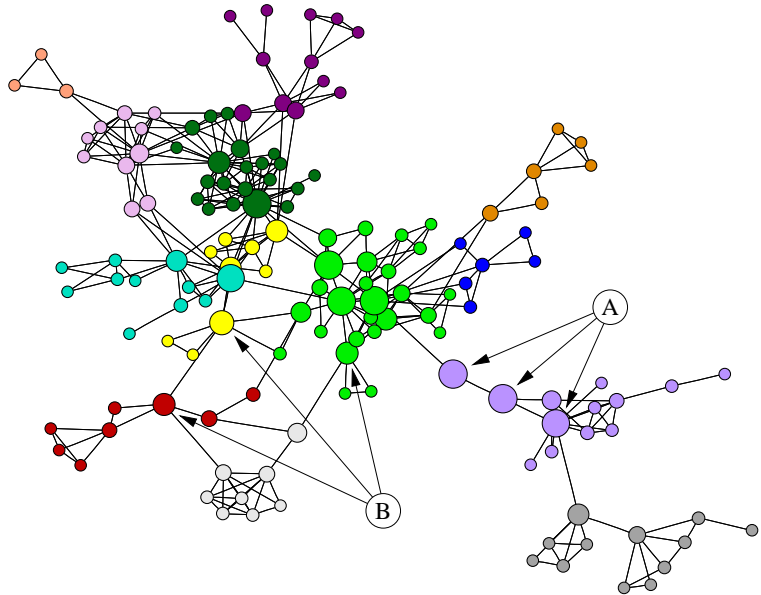}}
\end{center}
\caption{The largest component of the coauthorship network of scientists
working on networks from Newman and Park~\citey{NP03}.  Size of vertices
represents their score on the random-walk betweenness measure developed
here.  Vertices on a single path from one part of the network to another,
such as those labeled~``A,'' get a high score.  So however do those
labeled~``B,'' even though they lie on one of several paths between
different parts of the network.  The colors of the vertices represent one
possible clustering of the network, according to the method of Girvan and
Newman~\citey{GN02}, and are included as a guide to the eye.}
\label{collab}
\end{figure}

As a second example, we show in Fig.~\ref{collab} the largest component of
a coauthorship network taken from the study by Newman and
Park~\citey{NP03}.  The actors in this network are scientists, primarily in
applied mathematics and theoretical physics, who work on graph theory and
related mathematical studies of networks, and ties represent coauthorship
of papers.  As in Fig.~\ref{potterat}, the size of the vertices represents
their betweenness, calculated using our random-walk measure.  As we can see
there are a number of actors central to the groups in the network who have
high betweenness, although there are others who do not.  And there are less
central actors with high betweenness because they are the brokers who
establish connections between different groups (e.g.,~those labeled ``A''
in the figure).  But notice also that, where there are two (or more) paths
to an outlying group of vertices, those along all paths get a high score
(e.g.,~those labeled ``B''), since the random-walk betweenness counts all
paths and not just geodesic ones.

\section{Conclusions}
\label{concs}
Betweenness is a measure of network centrality that counts the paths
between vertex pairs on a network that pass through a given vertex.
Vertices with high betweenness lie on paths between many others and may
thus have some influence over the spread of information across the network.
One can define a variety of different betweenness measures, depending on
which paths one counts and how they are weighted.  The most widely used
measure, first proposed by Freeman~\citey{Freeman77}, counts only shortest
paths, and is thus appropriate to cases in which information flow is
entirely or mostly along such paths.  Flow betweenness
\citep{FBW91} counts all paths that carry information when a maximum
flow is pumped between each pair of vertices.  In many networks, however,
neither of these cases is realistic.  Both count only a small subset of
possible paths between vertices, and both assume some kind of optimality in
information transmission (shortest paths or maximum flow).

In this paper we have proposed a new betweenness measure that counts
essentially all paths between vertices (we exclude those that don't
actually lead from the designated source to the target), and which makes no
assumptions of optimality.  Our measure is based on random walks between
vertex pairs and asks, in essence, how often a given vertex will fall on a
random walk between another pair of vertices.  The measure is particularly
useful for finding vertices of high centrality that do not happen to lie on
geodesic paths or on the paths formed by maximum-flow cut-sets.  We have
shown that our betweenness can be calculated using matrix inversion methods
in time that scales as the cube of the number of vertices on a sparse
graph, making it computationally tractable for networks typical of current
sociological studies.

We have given a number of brief examples of the use of our betweenness
measure, including artificial examples illustrating cases in which it gives
substantially different results from previous measures, an example of how
it correlates with other measures in a network of sexual contacts, and two
applications to previously studied networks, Padgett's famous Florentine
families, and a network of collaborations between scientists.  We would be
delighted to see more, and more extensive, applications of our random-walk
betweenness measure in future studies.

\section*{Acknowledgments}
The author thanks Steven Borgatti and Michelle Girvan for useful comments
and conversations, and Richard Rothenberg and Stephen Muth for providing
the data for the example of Fig.~\ref{potterat}.  This work was funded in
part by the James S. McDonnell Foundation and by the National Science
Foundation under grant number DMS--0234188.

\end{document}